\documentclass[twocolumn]{aastex63}

\def\feh{$\mathrm{[Fe/H]}$}

\hypersetup{linkcolor=red,citecolor=blue,filecolor=green,urlcolor=magenta}

\begin{document}

\shorttitle{RR Lyrae \& Sextans II}
\shortauthors{Ngeow \& Bhardwaj}

\title{The Lack of RR Lyrae Variables in Ultrafaint Dwarf Galaxy Sextans II}

\correspondingauthor{Chow-Choong Ngeow}
\email{cngeow@astro.ncu.edu.tw}

\author[0000-0001-8771-7554]{Chow-Choong Ngeow}
\affil{Graduate Institute of Astronomy, National Central University, 300 Jhongda Road, 32001 Jhongli, Taiwan}

\author[0000-0001-6147-3360]{Anupam Bhardwaj}
\affil{Inter-University Center for Astronomy and Astrophysics (IUCAA), Post Bag 4, Ganeshkhind, Pune 411 007, India}

\begin{abstract}

  RR Lyrae in ultrafaint dwarf (UFD) galaxies, if found, would be valuable to constrain the distances of these UFD. In this work, we report our search of RR Lyrae in a recently discovered UFD -- Sextans II. Based on multiband ($griz$) time-series archival DECam imaging data, we do not find any RR Lyrae in Sextans II. On the contrary, the sparse multiband DECam light curves allowed recovery of a known foreground RR Lyrae, BB Sex, which happened to fall within the DECam images, and its pulsation period was correctly identified. Therefore, it is possible that Sextans II lacks RR Lyrae, similar to several other fainter UFDs (with $M_V \gtrsim -5.0$~mag) that do not have RR Lyrae. 

\end{abstract}


\section{Introduction}\label{sec1}

\begin{figure}
  \epsscale{1.1}
  \plotone{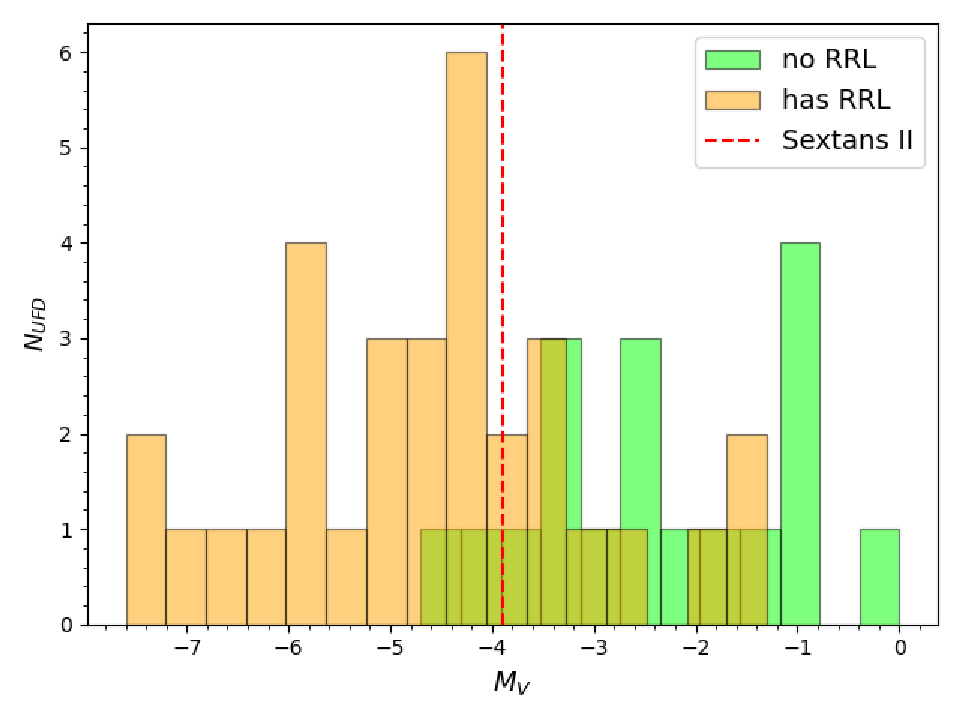}
  \caption{Comparison of the histograms for UFDs with and without detected RR Lyrae, based on the list compiled in \citet{ngeow2024b}. The vertical red dashed line marks the location of Sextans II.}   
  \label{fig_ufd}
\end{figure}

The search and study of the pristine ultrafaint dwarf (UFD) galaxies are one of the frontiers in astrophysics and of paramount importance for the near-field and/or small-scale cosmology. \citet{simon2019} provided a thorough review of UFDs, and defined those as dwarf galaxies with absolute $V$-band magnitude ($M_V$) fainter than $-7.7$~mag. Among the $\sim50$ UFDs currently known, $\sim64\%$ of them are hosting at least one RR Lyrae (see Figure \ref{fig_ufd}) -- an old population of pulsating stars that are an excellent distance indicator to trace accurate distance to their hosted UFD.

\begin{figure}
  \epsscale{1.1}
  \plotone{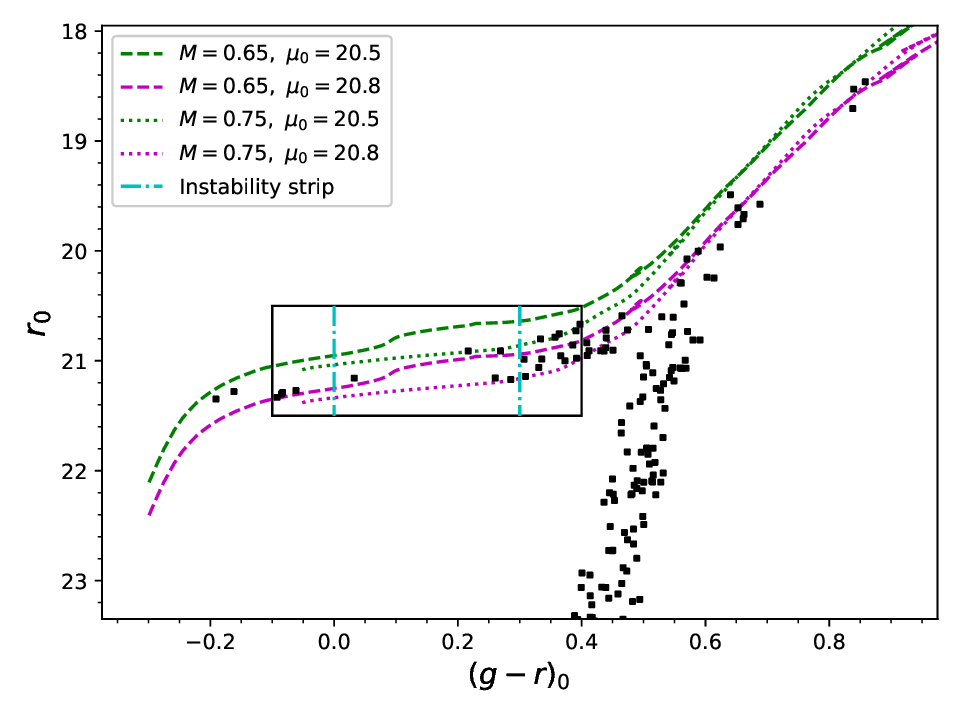}
  \caption{Color-magnitude diagram for Sextans II with possible members provided by M. Gatto (2024 -- private communication). The photometry for these possible members have been corrected for extinction \citep[for details see][]{gatto2024}. The dashed and dotted curves are the evolutionary tracks for two $\alpha$-enhanced horizontal branch mass models ($M$ in unit of Solar mass) at~\feh$=-2.2$~dex \citep{gatto2024,homma2023}, shifted with the distance moduli  ($\mu_0$) of $20.5$~mag \citep{homma2023} and $20.8$~mag \citep{gatto2024}. These models were taken from the {\tt BaSTI} library \citep{basti2021}. The vertical cyan dash-dotted lines marked the (rough) boundaries of the RR Lyrae instability strip. Stars within the rectangle were selected to search for RR Lyrae (see Section \ref{secrrl}).}   
  \label{fig_icmd}
\end{figure}

In recent years, new UFDs are routinely discovered from various deep imaging surveys. One of the newly identified UFDs is Sextans II, independently discovered by \citet[][using the Kilo-Degree Survey data]{gatto2024} and \citet[][using the Hyper Suprime-Cam Subaru Strategic Program Survey data]{homma2023}. Figure \ref{fig_icmd} presents the color-magnitude diagram (CMD) for possible members of Sextans II, where the list of possible members was provided by M. Gatto (2024 -- private communication). On top of this CMD, we have also included the horizontal branch (HB) evolutionary tracks at two representative mass models, taken from the {\tt BaSTI} \citep[a Bag of Stellar Tracks and Isochrones;][]{basti2021} stellar isochrones library. Since the RR Lyrae instability strip covers a color range, in $(g-r)$,\footnote{Note that both of the HB tracks and the instability strip, drawn in Figure \ref{fig_icmd}, are in the Sloan Digital Sky Survey (SDSS) photometric system, as both of the \citet{gatto2024} and \citet{homma2023} studies were done in the SDSS system.} from $\sim 0.0$~mag to $\sim 0.3$~mag \citep[][their Figure 6]{marconi2006}, and there are a few of the HB stars fall within this color range, it is possible that some of these HB stars are RR Lyrae. Furthermore, according to the empirical relation derived in \citet{mv2019}, Sextans II should have $2\pm1$ RR Lyrae when evaluating the relation at $M_V=-3.9$~mag \citep{gatto2024,homma2023}. 

In this work, we searched for RR Lyrae in Sextans II using the time-series archival Dark Energy Camera \citep[DECam,][]{decam2015} imaging data, due to its wide-field capability and it being mounted on a 4-meter class telescope (the Víctor M. Blanco Telescope). In Section \ref{secimg}, we describe the downloaded archival DECam images and the photometric calibration. We then search for possible RR Lyrae on the $\sim 30$ stars located within the rectangle box shown in Figure \ref{fig_icmd}. Note that we have enlarged this rectangle in color (by $\pm0.1$~mag on either side of the instability strip) to account for possible RR Lyrae located slightly beyond the theoretical instability strip of \citet{marconi2006}, due to the single-epoch observations of the data. We reported our search results in Section \ref{secrrl}, followed by discussion and conclusion in Section \ref{seclast}. 

\section{The DECam Archival Images} \label{secimg}

We searched and downloaded the $griz$-band multiepoch DECam images from the NOIRLab Astro Data Archive at the position of Sextans II. These DECam images were reduced using the DECam community pipeline \citep{valdes2014}, and with process types of {\tt InstCal}. Exposure time for these DECam images varied from 28~s to 250~s. The downloaded DECam images were in the multiextension fits (MEF) format, hence we split each of the MEF files into individual 61 CCD images using {\tt MissFITS} \citep{misfits2008}. We only retained the individual CCD images that contain (at least one) stars from the list of the possible members. Also, there are some stars that fall in the chip gap of the DECam focal plane. For the remaining stars, the minimum number of data points in all four filters is 1 (except a few stars do not have $i$-band data), and the median/maximum numbers of data points on the $griz$-band light curves are 12/16, 16/19, 9/12, and 13/17, respectively.

Photometric calibration for the reduced DECam images was done using the PS1 photometric catalog \citep{chambers2016,flewelling2020}; therefore, the calibrated magnitudes are in the PS1 AB magnitude system. For each image, we selected PS1 reference stars based on the criteria listed in the Appendix A of \citet[][applicable to $griz$ band]{ngeow2022}, and iteratively solved for the following equation:

\begin{eqnarray}
  m^{PS1}-m^{\mathrm{instr}} & = & ZP_m + \beta_m C^{PS1},
\end{eqnarray}

\noindent where $m=\{g,\ r,\ i,\ z\}$, and $C$ was set to $(g-r)$ and $(i-z)$ for the $gr$ band and $iz$ band, respectively. Point-spread function (PSF) photometry for the instrumental magnitudes, $m^{\mathrm{instr}}$, on both of the PS1 reference stars and the stars in the master stellar catalog were measured using the {\tt PSFEx} \citep{bertin2011} and {\tt Source-Extractor} \citep{bertin1996} software packages. {\tt Source-Extractor} was also used for source detection, as well as for cross-matching to the PS1 reference stars catalog and the master stellar catalog using its built-in association functionality.

\section{Searching for RR Lyrae Candidates} \label{secrrl}

\begin{figure}
  \epsscale{1.1}
  \plotone{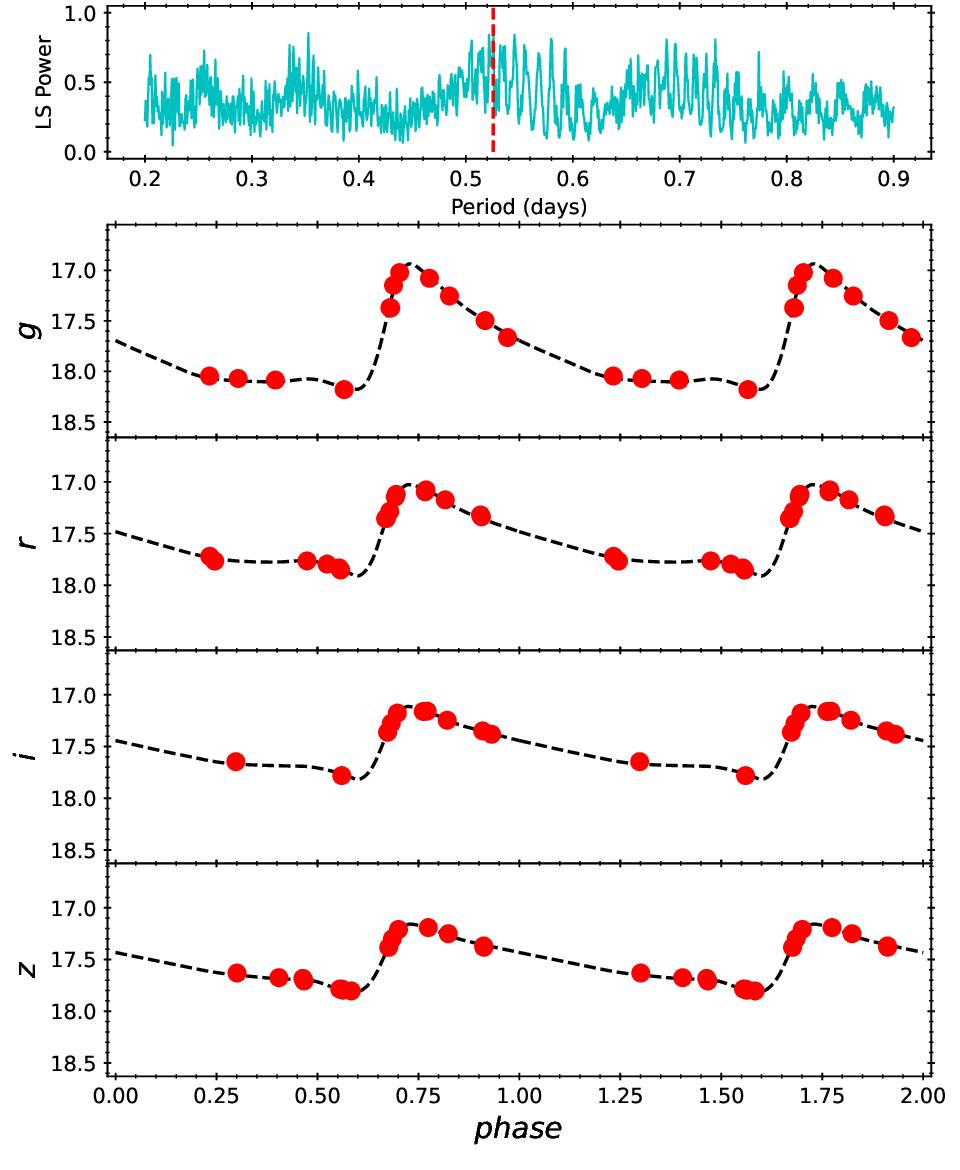}
  \caption{{\it Top Panel:} the multiband Lomb-Scargle (LS) periodogram for BB Sex based on the DECam light-curve data, the vertical dashed line indicates the best period found by {\tt gatspy}. {\it Bottom Panels:} folded $griz$-band DECam light curves based on the best-found period of $P=0.52563$~days. The black dashed curves are the best-fit template light curves adopted from \citet{sesar2010}. Note that the error bars are smaller than the size of the symbols.}
  \label{fig_bbsex}
\end{figure}

\begin{figure*}
  \epsscale{1.1}
  \gridline{\fig{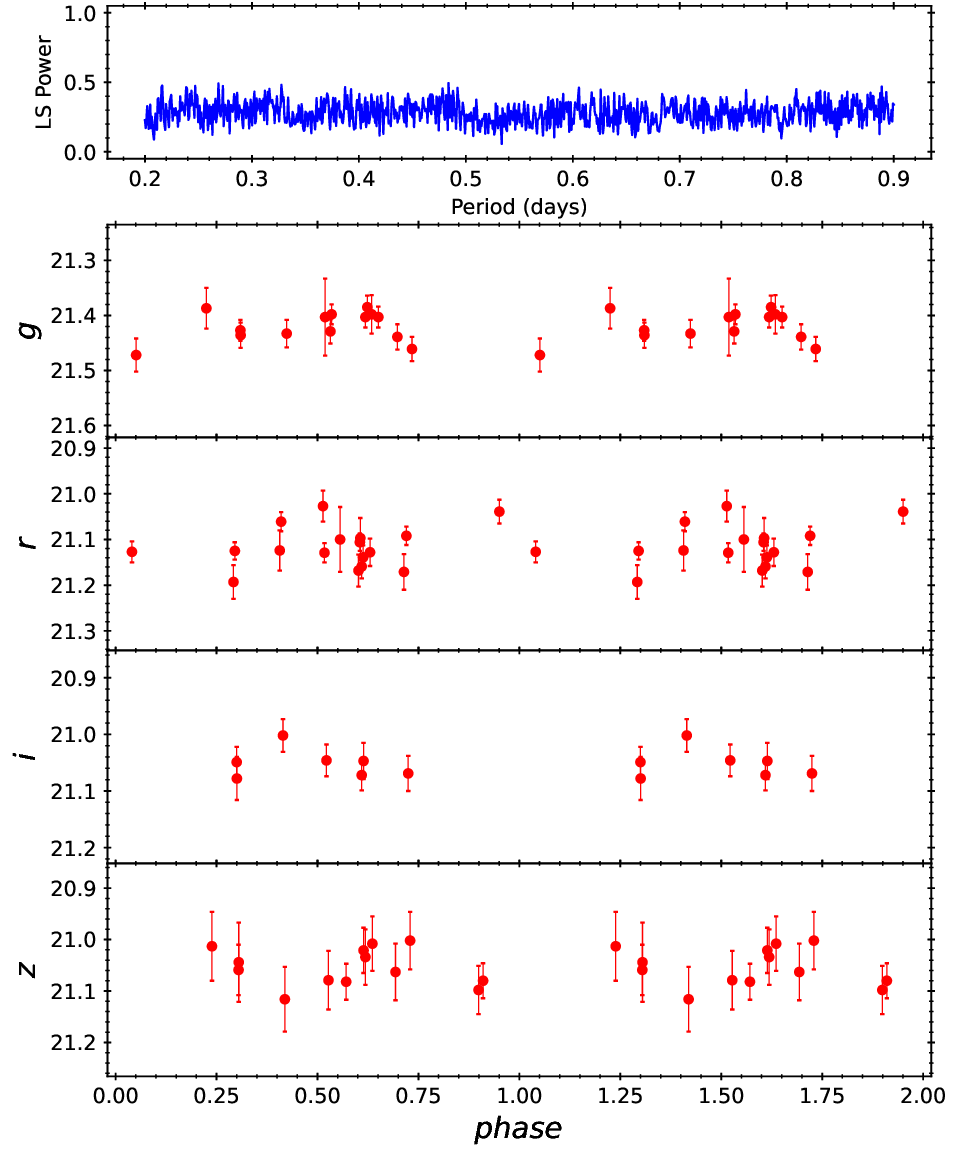}{0.3\textwidth}{(a) $P=0.32506$~days}
    \fig{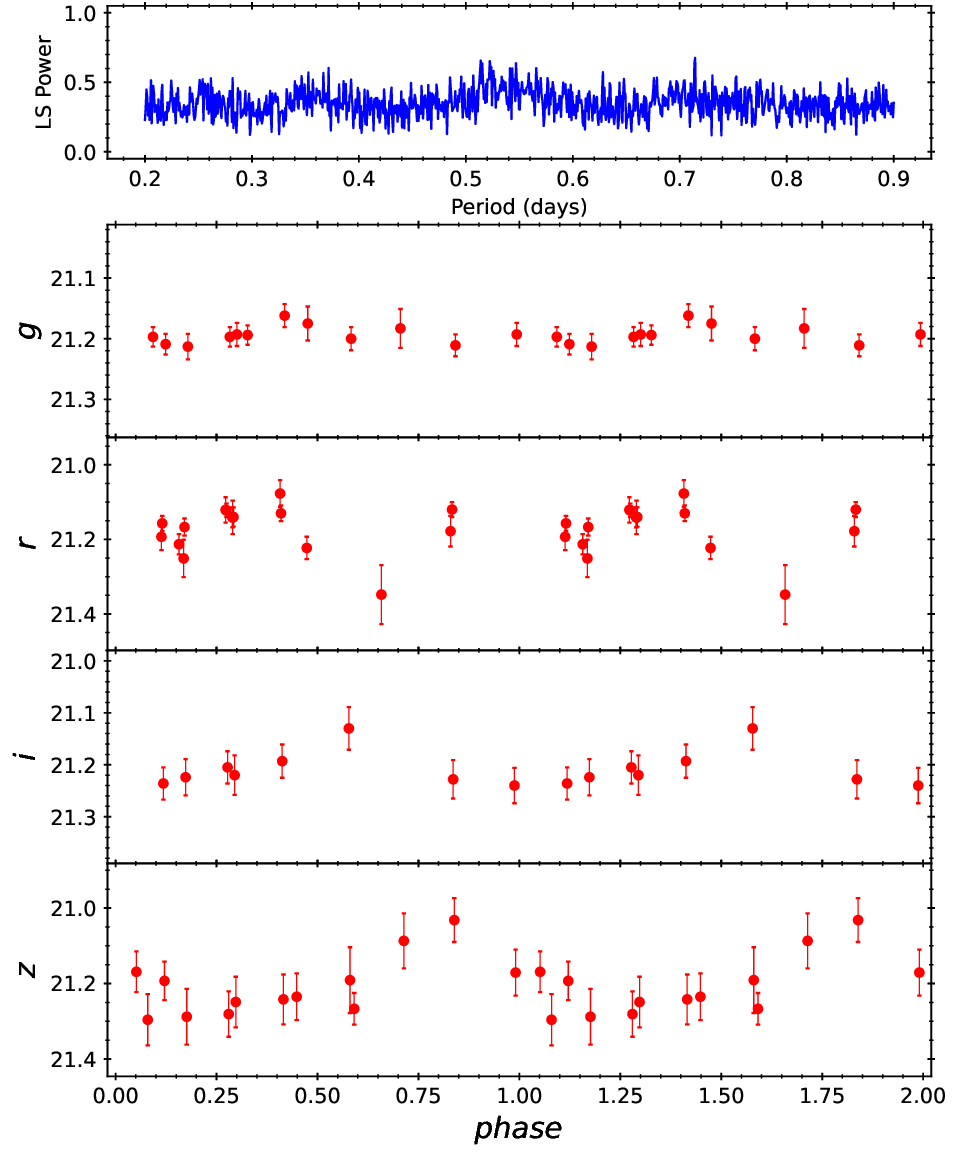}{0.3\textwidth}{(b) $P=0.53399$~days}
    \fig{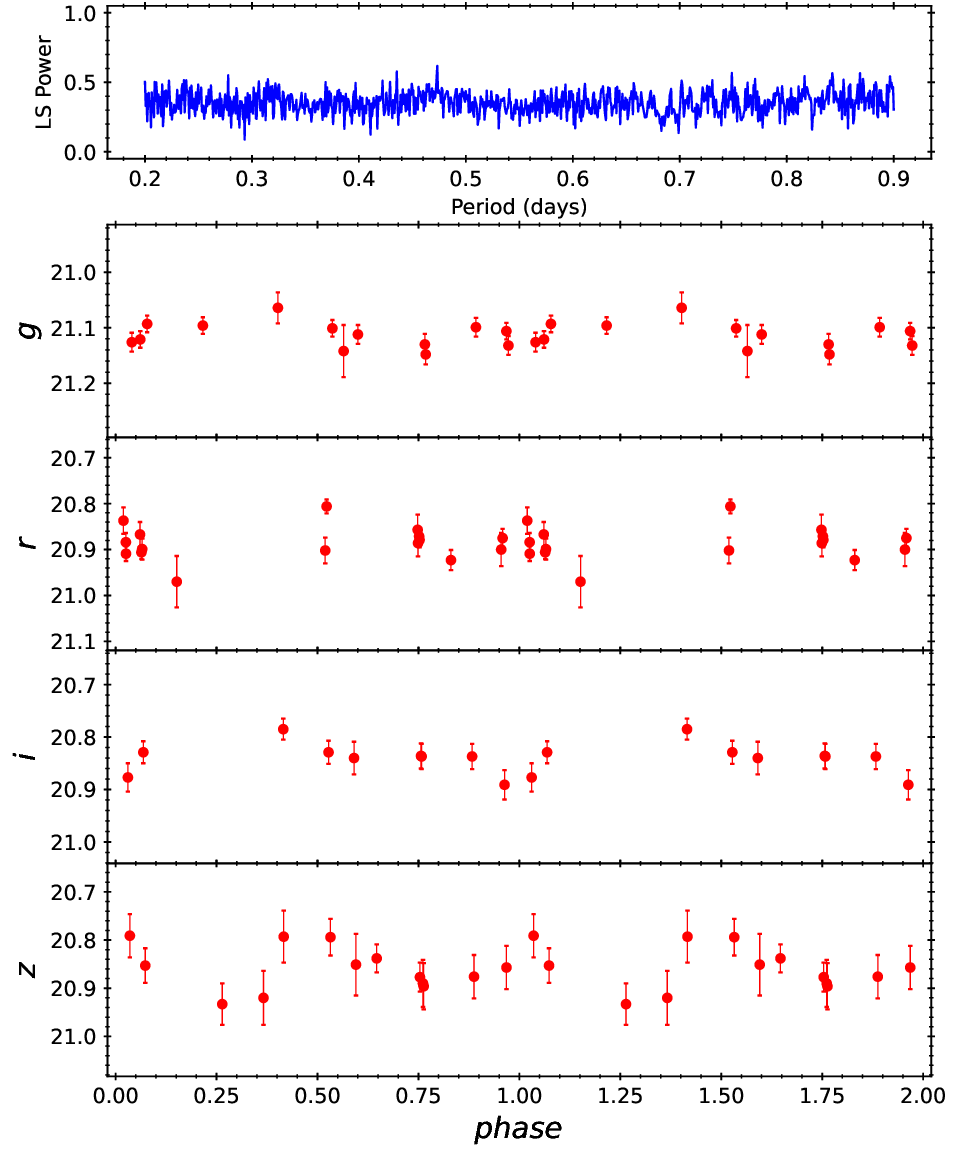}{0.3\textwidth}{(c) $P=0.31355$~days}}
  \gridline{\fig{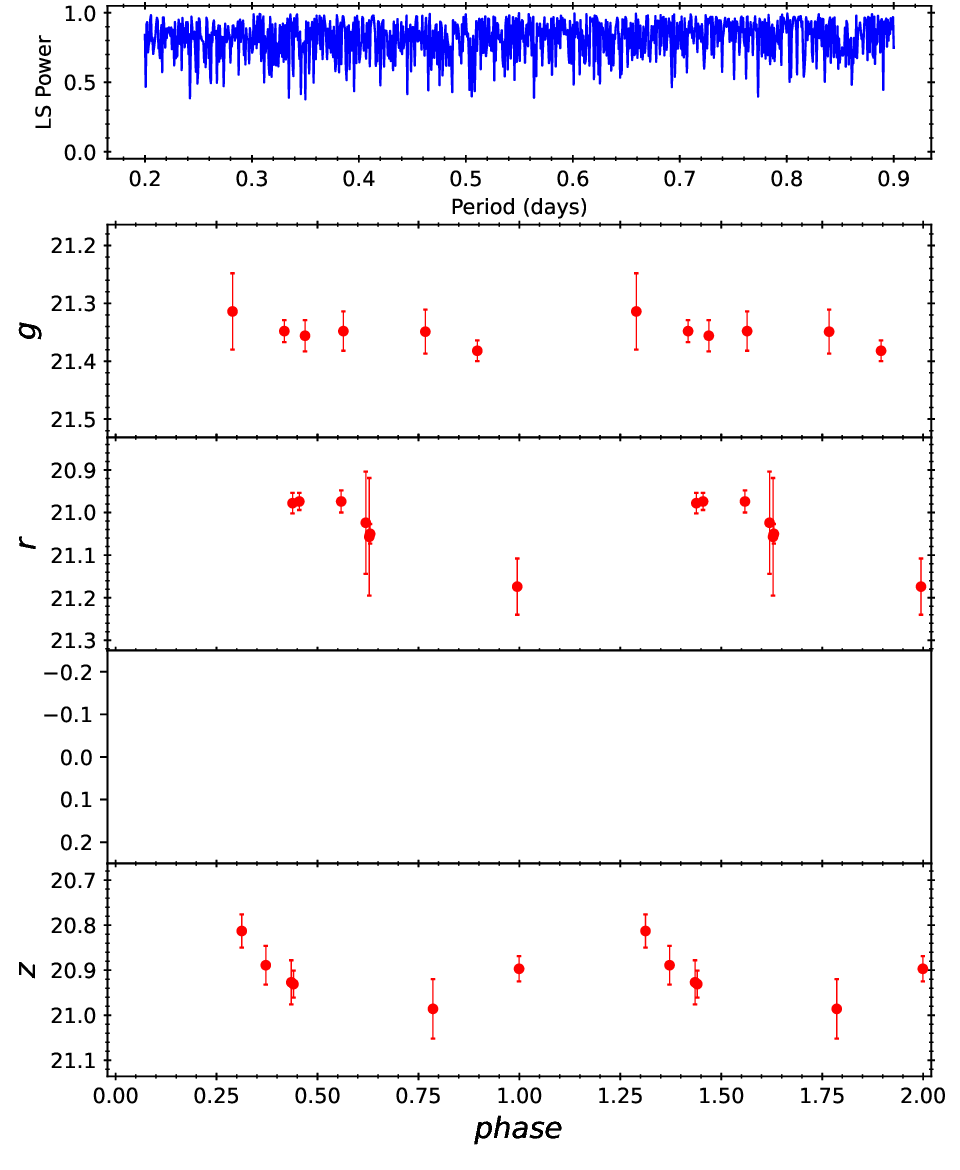}{0.3\textwidth}{(d) $P=0.23814$~days}
    \fig{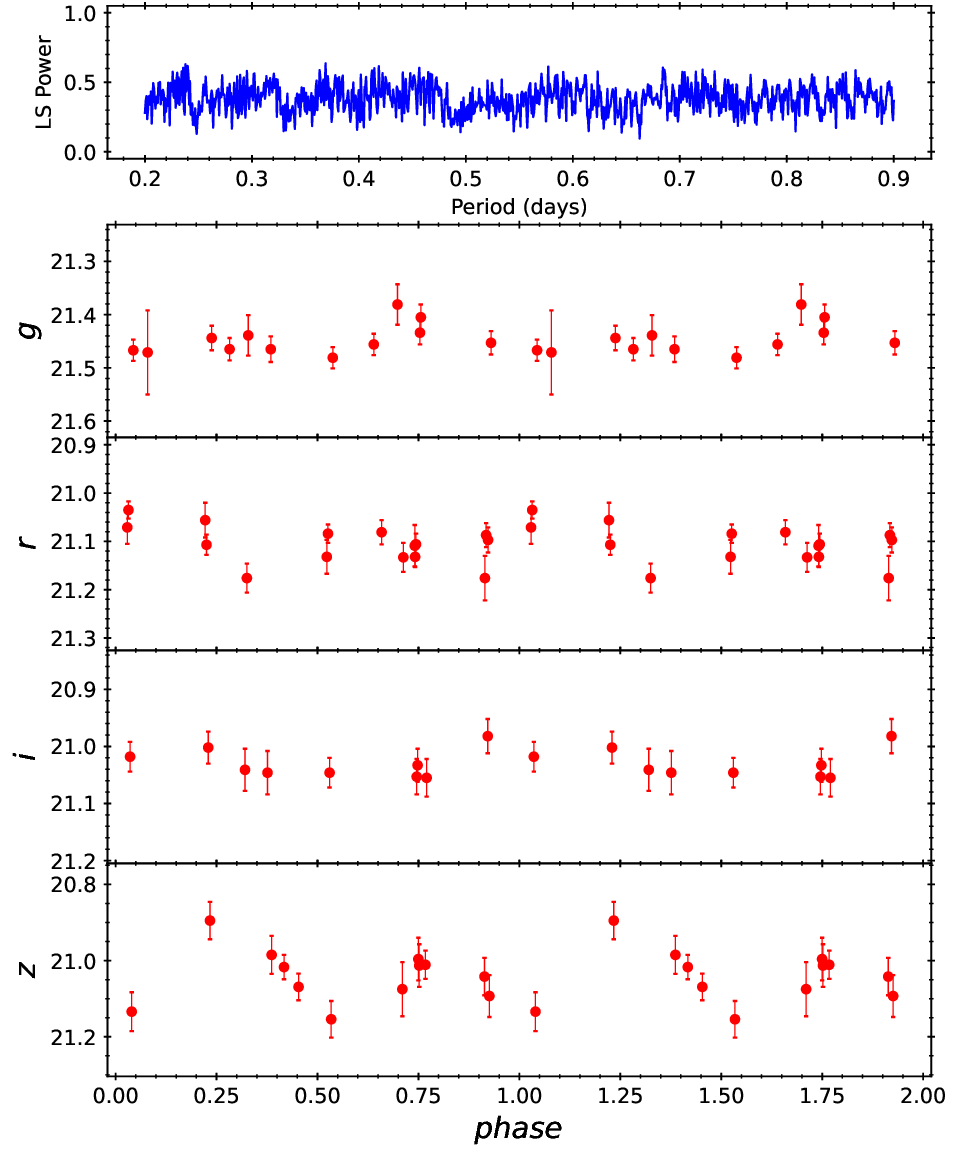}{0.3\textwidth}{(e) $P=0.37331$~days}
    \fig{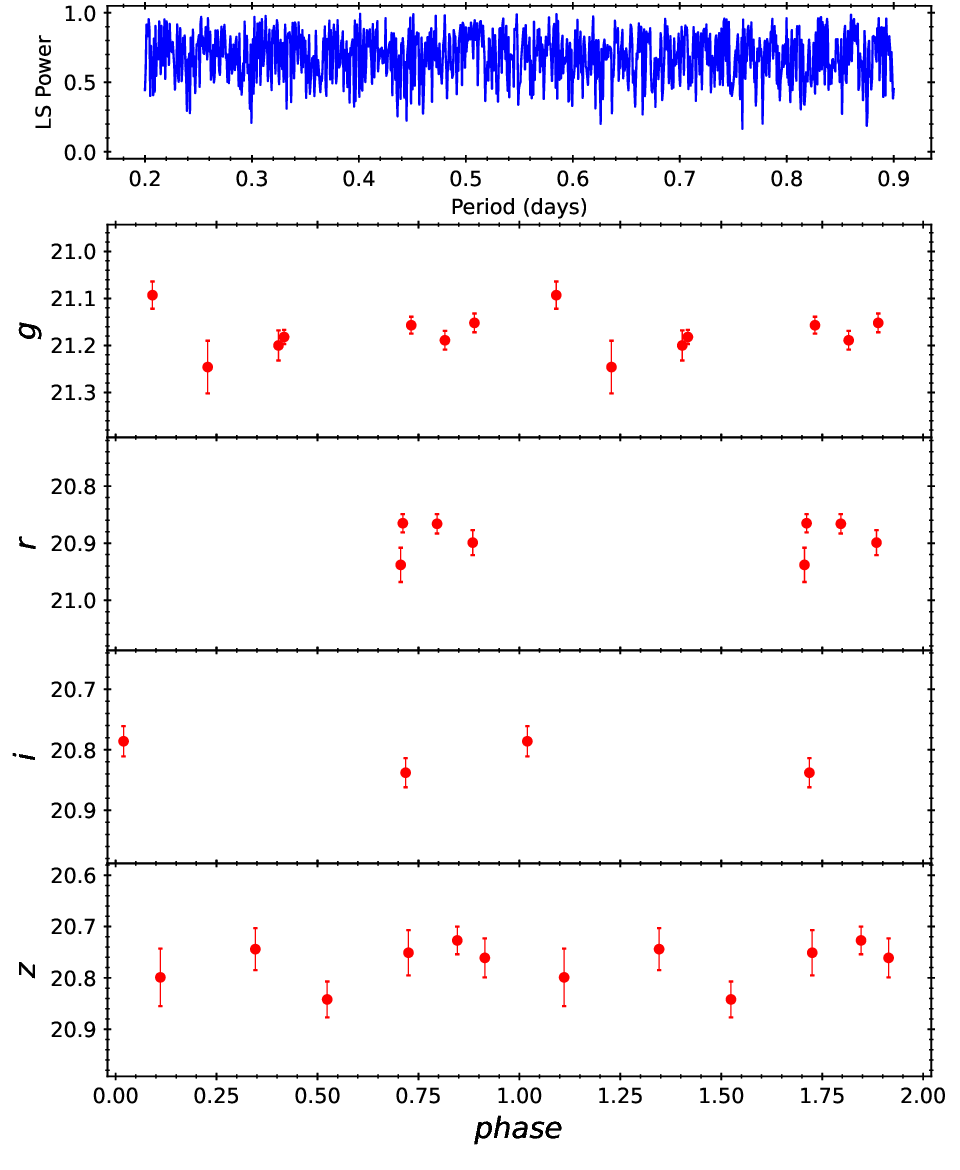}{0.3\textwidth}{(f) $P=0.23019$~days}}
  \caption{Similar to Figure \ref{fig_bbsex}, but for the six stars labeled in Figure \ref{fig_zoom}. The period $P$ referred to as the best period returned from {\tt gatspy}, which might not be the true period.}
  \label{fig_3star}
\end{figure*}

We ran the multiband Lomb-Scargle (LS) periodogram, implemented in the {\tt gatspy} Python package \citep{vdp2015}, to search for the possible periods among the $\sim 30$ stars located within the rectangle region shown in Figure \ref{fig_icmd}. The range of the searched period was set to be between 0.2 and 0.9~days (appropriate for RR Lyrae). The best-found periods returned from {\tt gatspy} (note that these periods might not correspond to the true periods) were then used to fold the $griz$-band light curves. 

\begin{figure}
  \epsscale{1.1}
  \plotone{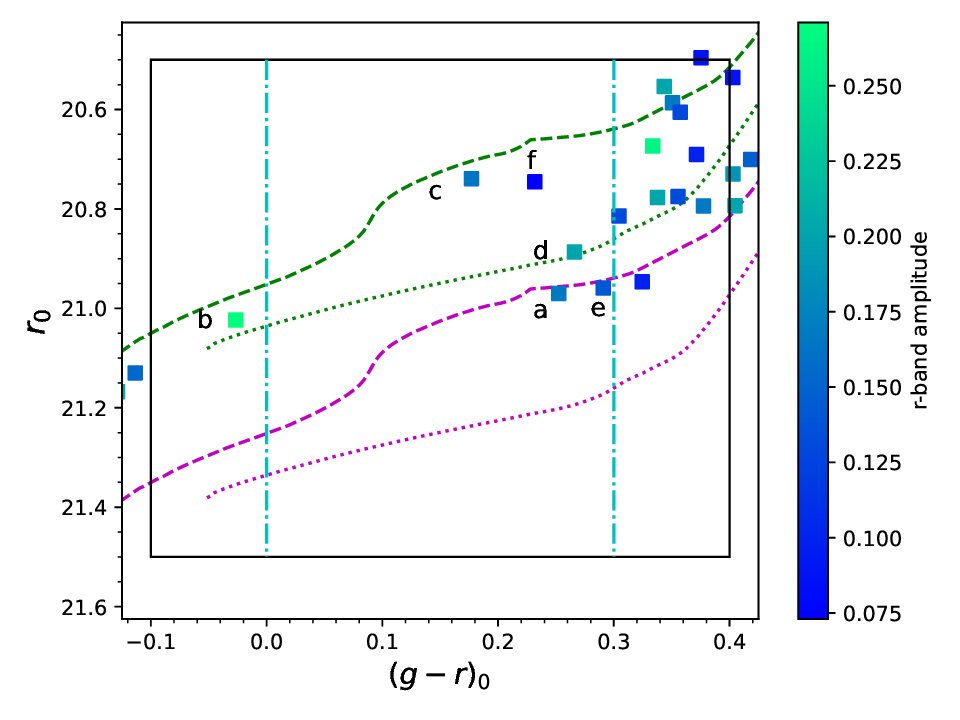}
  \caption{Zoomed-in version of Figure \ref{fig_icmd}. The color bar represents the $r$-band amplitudes, defined as the numerical difference between the maximum and minimum magnitudes of the observed data points. Note that the $gr$-band magnitudes are time-averaged values derived from the DECam light curves. Folded light curves for the six labeled stars, similar to Figure \ref{fig_bbsex}, are shown in Figure \ref{fig_3star}.}
  \label{fig_zoom}
\end{figure}

A known foreground RR Lyrae, BB Sex, was also located along the line of sight toward Sextans II, providing an opportunity to test our simple RR Lyrae search methods on the sparse multiband DECam light curves. The determined period for this RR Lyrae, using the DECam light curves, is $P(\mathrm{BB\ Sex}) = 0.52563$~days, which is almost identical to the values given in the Variable Stars Index \citep[VSX;][]{vsx} database: $P_{VSX}(\mathrm{BB\ Sex}) = 0.52562$~days. Lower panels of Figure \ref{fig_bbsex} present the folded DECam light curves for this foreground RR Lyrae, proving that potential RR Lyrae could be identified from the DECam multiband light curves.

Nevertheless, after visually inspecting the folded light curves for these $\sim30$ stars, we concluded that none of them are RR Lyrae, as those do not exhibit clear periodic and saw-tooth-shaped light curves typical for RR Lyrae. Example light curves for the six stars are shown in Figure \ref{fig_3star}. Furthermore, RR Lyrae are large-amplitude variable stars, with the largest amplitudes in the bluest filters (i.e. $gr$ band), and decrease toward the redder filters. Figure \ref{fig_zoom} shows the zoomed-in version of the CMD, showing color-coded $\sim 30$ stars that we searched for RR Lyrae with their observed $r$-band amplitudes (because the $r$ band has the most number of data points among the four filters). Almost all of them have a small $r$-band amplitude. It is true that some of them might be genuine variable stars; however, given our photometric precision and the number of epochs, a detailed investigation of them is out of the scope of this work.

\begin{figure}
  \epsscale{1.1}
  \plotone{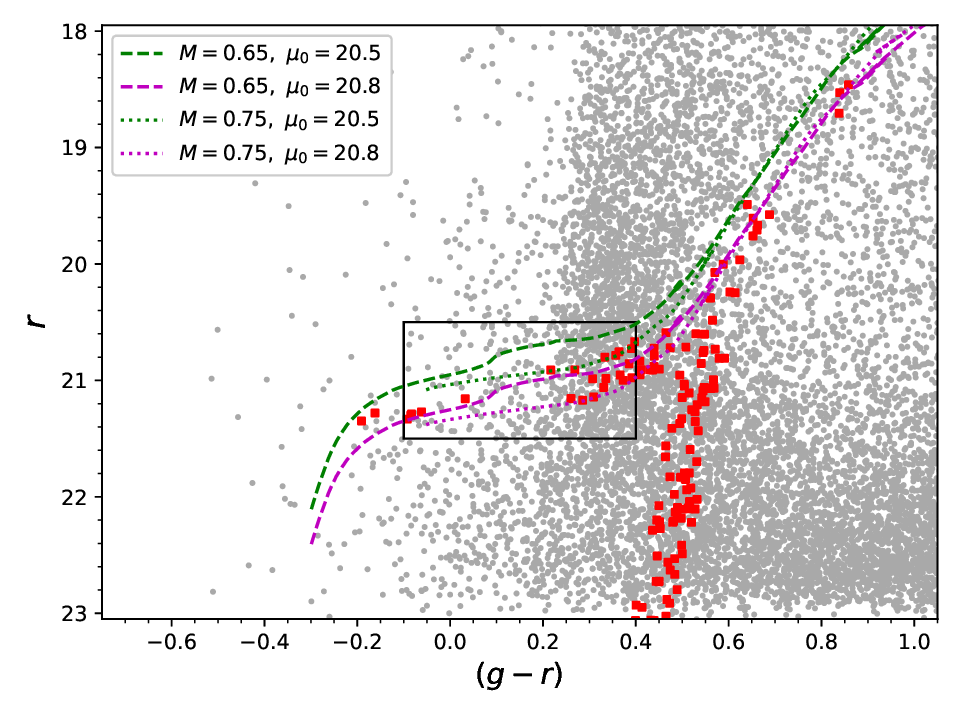}
  \caption{CMD for stars in the vicinity of Sextans II, including possible members (red squares) and additional stars queried from the SDSS DR16 (gray points; photometry for all of these stars are in the SDSS system). We only search the potential RR Lyrae for SDSS stars located inside the rectangle, which is the same as in Figure \ref{fig_icmd}. The HB tracks are the same as in Figure \ref{fig_icmd}.}
  \label{fig_lcmd}
\end{figure}

To account for the possibility that the list of possible members missed the RR Lyre in Sextans II, we expanded the search area using a radius of $\sim 15r_h$ (where $r_h$ is the half-light radius of Sextans II). Figure \ref{fig_lcmd} presents the CMD for stars located within the search area, including stars queried from the Sload Digital Sky Survey (SDSS) Data Release 16 \citep[DR16,][]{sdss2020}. Out of the $\sim 730$ SDSS stars located within the rectangle shown in Figure \ref{fig_lcmd}, $\sim 70$ of them have good DECam light-curve quality and exhibit variability for a further inspection. However, after visually inspecting their light curves, none of them turned out to be a RR Lyrae.

\section{Discussion and Conclusion} \label{seclast}

\begin{figure*}
  \epsscale{1.1}
  \plottwo{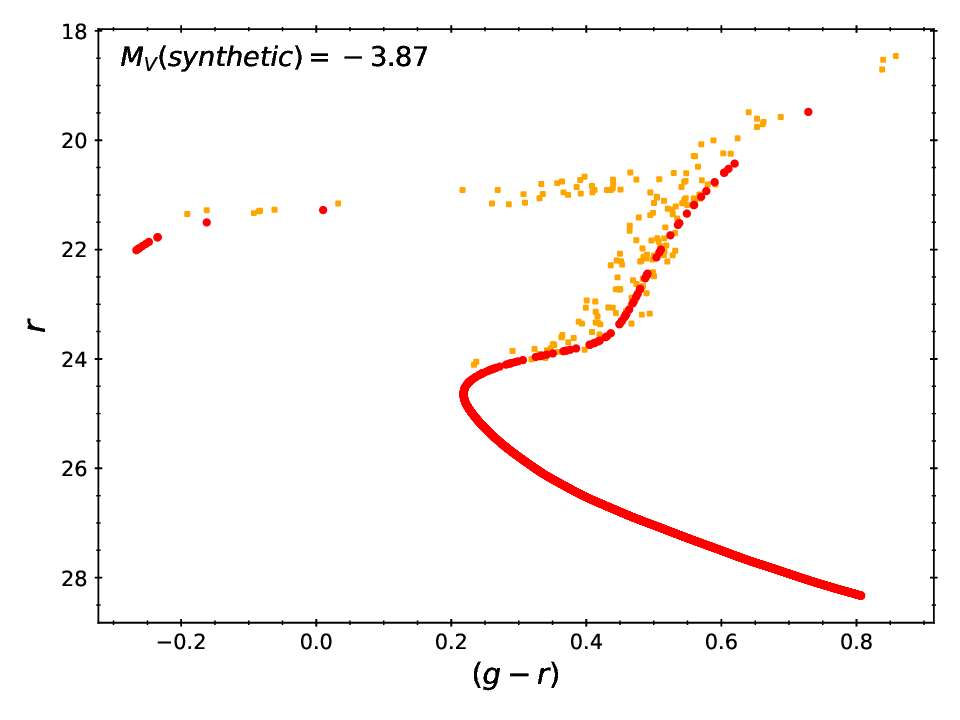}{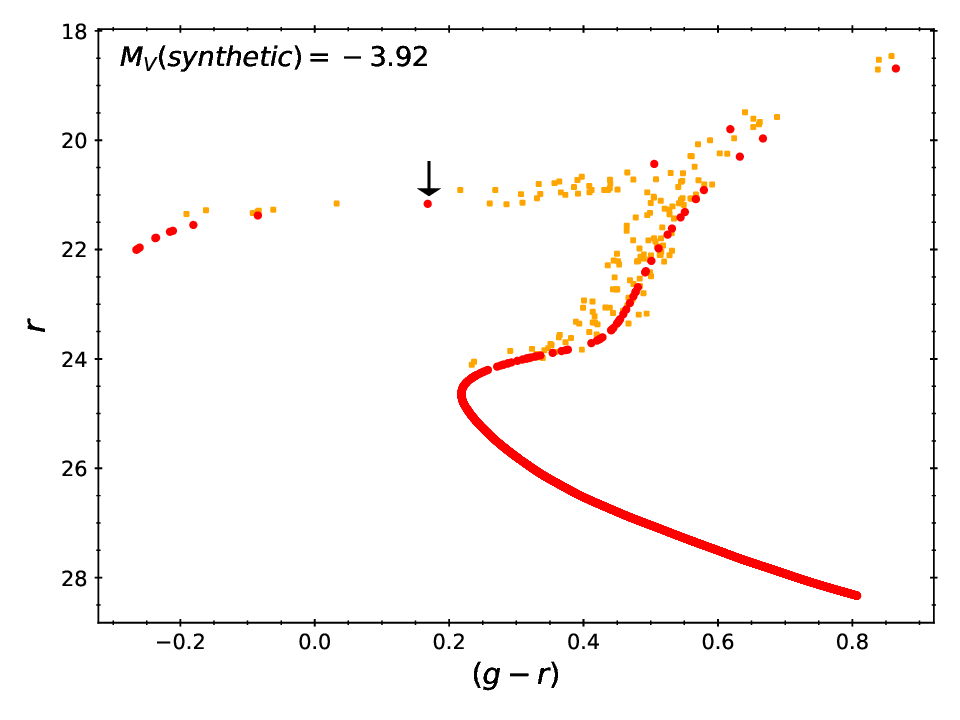}
  \caption{Comparisons of the two $\alpha$-enhanced synthetic CMD (red points, vertically shifted using $\mu_0=20.8$~mag), generated from using the {\tt BaSTI} online tool \citep{basti2021}, to the possible members of Sextans II (orange points). The input parameters for both of the synthetic CMD are: age of 13 Gyrs, \feh$=-2.2$~dex, no binary fraction, and using the \citet{kroupa1993} initial mass function. The integrated $gr$-band magnitudes of the synthetic CMD were converted to $V$ band using the transformation given in \citet{jester2005}, and the resulting $M_V$ are given in the top-left corner of each panel. The vertical arrow on the right panel indicates the synthetic RR Lyrae identified from the synthetic CMD by {\tt BaSTI}. }
  \label{fig_syn}
\end{figure*}

In this work, we searched for possible RR Lyre in Sextans II, a newly discovered UFD, using the time-series DECam archival data and found none. However, it is worth pointing out that the archival DECam data\footnote{We have also searched for the Canada France Hawaii Telescope (CFHT) MegaCam public archive, but no data were found at the location of Sextans II when this work was conducted.} are not optimal for the search of RR Lyrae, especially in terms of the number of epochs and the heterogeneous exposure time. A dedicated time-series observation aimed at searching for RR Lyrae in Sextans II, using a wide-field imager on 4-8 meter-class telescopes, is warranted.

At $M_V=-3.9$~mag, Sextans II falls within the range of UFD with and without RR Lyrae (Figure \ref{fig_ufd}). Therefore, it is also entirely possible that Sextans II intrinsically does not have any RR Lyrae \citep[as similar to Willman 1;][]{siegel2008}. Using {\tt BaSTI}, together with parameters adopted from \citet{gatto2024}, we generated two synthetic CMDs with $M_V$ bracketing $-3.9$~mag. Input parameters for both of the synthetic CMDs are almost identical, except for the number of input stars (hence determining the $M_V$). These two synthetic CMDs are displayed in Figure \ref{fig_syn}. The {\tt BaSTI} tool also returned the number of RR Lyrae if there are synthetic stars fall inside the instability strip of the Hertzsprung-Russell diagram \citep[instead of the CMD; see Appendix in][]{basti2021}. The synthetic CMD in the left panel of Figure \ref{fig_syn}, with $M_V=-3.87$~mag, does not have any RR Lyrae. In contrast, the synthetic CMD in the right panel of Figure \ref{fig_syn}, with a slightly brighter $M_V$ of $-3.92$~mag, contained one synthetic RR Lyrae (marked as the arrow in Figure \ref{fig_syn}). This shows that, at $M_V\sim -3.9$~mag, the presence of RR Lyrae is sensitive to the true number of member stars of the UFD, and Sextans II seems to fall on the borderline of presence/absence of RR Lyrae. 

Among the 38 UFDs with $M_V \geq -5.0$~mag, 20 host at least one RR Lyrae, giving a fraction of $52.6\%$. After including the result of Sextans II, this fraction drops slightly to $51.3\%$. However, new UFDs are yet to be discovered, and not all of the currently known UFDs have been searched for RR Lyrae. Therefore, either detection or nondetection of RR Lyrae in a newly discovered UFD will be valuable to constrain the true fraction of UFDs hosting RR Lyrae.

\acknowledgments

We thank the useful discussions and comments from an anonymous referee to improve the manuscript. We sincerely thank M. Gatto for sharing the membership list of Sextans II. We also thank S. Cassisi for the assistance on using the {\tt BaSTI} web tool. We are thankful for funding from the National Science and Technology Council (Taiwan) under the grant 113-2112-M-008-028. This research has made use of the SIMBAD database and the VizieR catalogue access tool, operated at CDS, Strasbourg, France. This research made use of Astropy,\footnote{\url{http://www.astropy.org}} a community-developed core Python package for Astronomy \citep{astropy2013, astropy2018, astropy2022}. This research has made use of the International Variable Star Index (VSX) database, operated at AAVSO, Cambridge, Massachusetts, USA.

The Pan-STARRS1 Surveys (PS1) and the PS1 public science archive have been made possible through contributions by the Institute for Astronomy, the University of Hawaii, the Pan-STARRS Project Office, the Max-Planck Society and its participating institutes, the Max Planck Institute for Astronomy, Heidelberg and the Max Planck Institute for Extraterrestrial Physics, Garching, The Johns Hopkins University, Durham University, the University of Edinburgh, the Queen's University Belfast, the Harvard-Smithsonian Center for Astrophysics, the Las Cumbres Observatory Global Telescope Network Incorporated, the National Central University of Taiwan, the Space Telescope Science Institute, the National Aeronautics and Space Administration under Grant No. NNX08AR22G issued through the Planetary Science Division of the NASA Science Mission Directorate, the National Science Foundation Grant No. AST-1238877, the University of Maryland, Eotvos Lorand University (ELTE), the Los Alamos National Laboratory, and the Gordon and Betty Moore Foundation.

Funding for the Sloan Digital Sky Survey IV has been provided by the Alfred P. Sloan Foundation, the U.S. Department of Energy Office of Science, and the Participating Institutions. SDSS acknowledges support and resources from the Center for High-Performance Computing at the University of Utah. The SDSS web site is \url{www.sdss4.org}.

SDSS is managed by the Astrophysical Research Consortium for the Participating Institutions of the SDSS Collaboration including the Brazilian Participation Group, the Carnegie Institution for Science, Carnegie Mellon University, Center for Astrophysics | Harvard \& Smithsonian (CfA), the Chilean Participation Group, the French Participation Group, Instituto de Astrofísica de Canarias, The Johns Hopkins University, Kavli Institute for the Physics and Mathematics of the Universe (IPMU) / University of Tokyo, the Korean Participation Group, Lawrence Berkeley National Laboratory, Leibniz Institut für Astrophysik Potsdam (AIP), Max-Planck-Institut fur Astronomie (MPIA Heidelberg), Max-Planck-Institut für Astrophysik (MPA Garching), Max-Planck-Institut für Extraterrestrische Physik (MPE), National Astronomical Observatories of China, New Mexico State University, New York University, University of Notre Dame, Observatório Nacional / MCTI, The Ohio State University, Pennsylvania State University, Shanghai Astronomical Observatory, United Kingdom Participation Group, Universidad Nacional Autónoma de México, University of Arizona, University of Colorado Boulder, University of Oxford, University of Portsmouth, University of Utah, University of Virginia, University of Washington, University of Wisconsin, Vanderbilt University, and Yale University.

This research draws upon DECam data as distributed by the Astro Data Archive at NSF NOIRLab. NOIRLab is managed by the Association of Universities for Research in Astronomy (AURA) under a cooperative agreement with the U.S. National Science Foundation.

This project used data obtained with the Dark Energy Camera (DECam), which was constructed by the Dark Energy Survey (DES) collaboration. Funding for the DES Projects has been provided by the US Department of Energy, the US National Science Foundation, the Ministry of Science and Education of Spain, the Science and Technology Facilities Council of the United Kingdom, the Higher Education Funding Council for England, the National Center for Supercomputing Applications at the University of Illinois at Urbana-Champaign, the Kavli Institute for Cosmological Physics at the University of Chicago, Center for Cosmology and Astro-Particle Physics at the Ohio State University, the Mitchell Institute for Fundamental Physics and Astronomy at Texas A\&M University, Financiadora de Estudos e Projetos, Fundação Carlos Chagas Filho de Amparo à Pesquisa do Estado do Rio de Janeiro, Conselho Nacional de Desenvolvimento Científico e Tecnológico and the Ministério da Ciência, Tecnologia e Inovação, the Deutsche Forschungsgemeinschaft and the Collaborating Institutions in the Dark Energy Survey.

The Collaborating Institutions are Argonne National Laboratory, the University of California at Santa Cruz, the University of Cambridge, Centro de Investigaciones Enérgeticas, Medioambientales y Tecnológicas–Madrid, the University of Chicago, University College London, the DES-Brazil Consortium, the University of Edinburgh, the Eidgenössische Technische Hochschule (ETH) Zürich, Fermi National Accelerator Laboratory, the University of Illinois at Urbana-Champaign, the Institut de Ciències de l’Espai (IEEC/CSIC), the Institut de Física d’Altes Energies, Lawrence Berkeley National Laboratory, the Ludwig-Maximilians Universität München and the associated Excellence Cluster Universe, the University of Michigan, NSF NOIRLab, the University of Nottingham, the Ohio State University, the OzDES Membership Consortium, the University of Pennsylvania, the University of Portsmouth, SLAC National Accelerator Laboratory, Stanford University, the University of Sussex, and Texas A\&M University.

Based in part on observations at Cerro Tololo Inter-American Observatory, National Optical Astronomy Observatory, which is operated by the Association of Universities for Research in Astronomy (AURA) under a cooperative agreement with the National Science Foundation.

\facility{Blanco (DECam)}

\software{{\tt astropy} \citep{astropy2013,astropy2018,astropy2022}, {\tt gatspy} \citep{vdp2015},  {\tt MissFITS} \citep{misfits2008}, {\tt PSFEx} \citep{bertin2011}, {\tt Source-Extractor} \citep{bertin1996}}



\begin{thebibliography}{} 

\bibitem[Ahumada et al.(2020)]{sdss2020} Ahumada, R., Allende Prieto, C., Almeida, A., et al.\ 2020, \apjs, 249, 3

\bibitem[Astropy Collaboration et al.(2013)]{astropy2013} Astropy Collaboration, Robitaille, T.~P., Tollerud, E.~J., et al.\ 2013, \aap, 558, A33

\bibitem[Astropy Collaboration et al.(2018)]{astropy2018} Astropy Collaboration, Price-Whelan, A.~M., Sip{\H{o}}cz, B.~M., et al.\ 2018, \aj, 156, 123

\bibitem[Astropy Collaboration et al.(2022)]{astropy2022} Astropy Collaboration, Price-Whelan, A.~M., Lim, P.~L., et al.\ 2022, \apj, 935, 167

\bibitem[Bertin(2011)]{bertin2011} Bertin, E.\ 2011, in ASP Conf. Ser. 442, Astronomical Data Analysis Software and Systems XX, ed. I. N. Evans et al. (San Francisco, CA: ASP), 435

\bibitem[Bertin \& Arnouts(1996)]{bertin1996} Bertin, E. \& Arnouts, S.\ 1996, \aaps, 117, 393

\bibitem[Chambers et al.(2016)]{chambers2016} Chambers, K.~C., Magnier, E.~A., Metcalfe, N., et al.\ 2016, arXiv:1612.05560
    
\bibitem[Flaugher et al.(2015)]{decam2015} Flaugher, B., Diehl, H.~T., Honscheid, K., et al.\ 2015, \aj, 150, 150

\bibitem[Flewelling et al.(2020)]{flewelling2020} Flewelling, H.~A., Magnier, E.~A., Chambers, K.~C., et al.\ 2020, \apjs, 251, 7

\bibitem[Gatto et al.(2024)]{gatto2024} Gatto, M., Bellazzini, M., Tortora, C., et al.\ 2024, \aap, 681, L13

    
\bibitem[Homma et al.(2024)]{homma2023} Homma, D., Chiba, M., Komiyama, Y., et al.\ 2024, \pasj, 76, 733

\bibitem[Jester et al.(2005)]{jester2005} Jester, S., Schneider, D.~P., Richards, G.~T., et al.\ 2005, \aj, 130, 873
  
\bibitem[Kroupa et al.(1993)]{kroupa1993} Kroupa, P., Tout, C.~A., \& Gilmore, G.\ 1993, \mnras, 262, 545
  
\bibitem[Marconi et al.(2006)]{marconi2006} Marconi, M., Cignoni, M., Di Criscienzo, M., et al.\ 2006, \mnras, 371, 1503
  
\bibitem[Marmo \& Bertin(2008)]{misfits2008} Marmo, C. \& Bertin, E.\ 2008, ASP Conf. Series 394, Astronomical Data Analysis Software and Systems XVII, ed. R. W. Argyle, P. S. Bunclark, and J. R. Lewis (San Francisco, CA: ASP), 619
  
\bibitem[Mart{\'\i}nez-V{\'a}zquez et al.(2019)]{mv2019} Mart{\'\i}nez-V{\'a}zquez, C.~E., Vivas, A.~K., Gurevich, M., et al.\ 2019, \mnras, 490, 2183
  
\bibitem[Ngeow(2022)]{ngeow2022} Ngeow, C.-C.\ 2022, \aj, 164, 45
  
\bibitem[Ngeow \& Bhardwaj(2024)]{ngeow2024b} Ngeow, C.-C. \& Bhardwaj, A.\ 2024, \aj, 168, 8

\bibitem[Pietrinferni et al.(2021)]{basti2021} Pietrinferni, A., Hidalgo, S., Cassisi, S., et al.\ 2021, \apj, 908, 102

\bibitem[Sesar et al.(2010)]{sesar2010} Sesar, B., Ivezi{\'c}, {\v{Z}}., Grammer, S.~H., et al.\ 2010, \apj, 708, 717

\bibitem[Siegel et al.(2008)]{siegel2008} Siegel, M.~H., Shetrone, M.~D., \& Irwin, M.\ 2008, \aj, 135, 2084
  
\bibitem[Simon(2019)]{simon2019} Simon, J.~D.\ 2019, \araa, 57, 375

\bibitem[Valdes et al.(2014)]{valdes2014} Valdes, F., Gruendl, R., \& DES Project\ 2014, ASP Conf. Series 485, Astronomical Data Analysis Software and Systems XXIII, ed. N. Manset and P. Forshay (San Francisco, CA: ASP), 379
  
\bibitem[VanderPlas \& Ivezi{\'c}(2015)]{vdp2015} VanderPlas, J.~T., \& Ivezi{\'c}, {\v{Z}}.\ 2015, \apj, 812, 18

\bibitem[Watson et al.(2006)]{vsx} Watson, C.~L., Henden, A.~A., \& Price, A.\ 2006, Society for Astronomical Sciences Annual Symposium, 25, 47
  
\end{thebibliography}
\end{document}